\documentclass{article}

\usepackage{arxiv}
\usepackage{lmodern}
\usepackage{graphicx}
\usepackage{amsmath}
\usepackage{bm}
\usepackage{enumitem}
\usepackage{float}
\usepackage{caption}
\usepackage{subcaption}
\usepackage{array}
\usepackage{amsfonts}
\usepackage[dvipsnames]{xcolor}
\usepackage{booktabs}
\usepackage{nomencl}
\makenomenclature
\usepackage{arydshln}
\usepackage{pdfpages}
\usepackage{placeins}
\usepackage{multirow}
\usepackage{scalerel}[2016/12/29]
\newcommand\Undersig{\scaleobj{0.75}{\Sigma}}

\usepackage{algorithm}
\usepackage[noend]{algpseudocode}
\usepackage{hyperref}

\begin{document}
\title{Context-Aware Generative Models for Prediction of Aircraft Ground Tracks}
%
%


\author{Nick Pepper\IEEEauthorrefmark{1}, George {De Ath}\IEEEauthorrefmark{2}, Marc Thomas\IEEEauthorrefmark{3}, Richard Everson\IEEEauthorrefmark{2} and Tim Dodwell\IEEEauthorrefmark{2}\thanks{\IEEEauthorrefmark{1} The Alan Turing Institute \IEEEauthorrefmark{2} University of Exeter \IEEEauthorrefmark{3} NATS. Manuscript received xxxx; revised xxxx. This work was supported by an EPSRC Prosperity Partnership between NATS, Turing, Exeter and Cambridge.}}

\author{
    Nick Pepper \\
	The Alan Turing Institute\\
	The British Library\\
	London, UK \\
	\texttt{npepper@turing.ac.uk} \\
    \And
    George {De Ath}  \\
	Department of Computer Science\\
	University of Exeter\\
	Exeter, UK \\
    \And
    Marc Thomas  \\
	NATS\\
	Whitely, Fareham, UK \\
    \And
    Richard Everson  \\
	Department of Computer Science\\
	University of Exeter\\
	Exeter, UK \\
    \And
    Tim Dodwell  \\
	Department of Engineering\\
	University of Exeter\\
	Exeter, UK \\
}

\maketitle
\begin{abstract}
Trajectory prediction (TP) plays an important role in supporting the decision-making of Air Traffic Controllers (ATCOs). Traditional TP methods are deterministic and physics-based, with parameters that are calibrated using aircraft surveillance data harvested across the world. These models are, therefore, agnostic to the intentions of the pilots and ATCOs, which can have a significant effect on the observed trajectory, particularly in the lateral plane. This work proposes a generative method for lateral TP, using probabilistic machine learning to model the effect of the epistemic uncertainty arising from the unknown effect of pilot behaviour and ATCO intentions. The models are trained to be specific to a particular sector, allowing local procedures such as coordinated entry and exit points to be modelled. A dataset comprising a week's worth of aircraft surveillance data, passing through a busy sector of the United Kingdom's upper airspace, was used to train and test the models. Specifically, a piecewise linear model was used as a functional, low-dimensional representation of the ground tracks, with its control points determined by a generative model conditioned on partial context. It was found that, of the investigated models, a Bayesian Neural Network using the Laplace approximation was able to generate the most plausible trajectories in order to emulate the flow of traffic through the sector.\looseness=-1

\end{abstract}


\section{Introduction}
\label{sec:intro}

Air Traffic Control (ATC) issues instructions to aircraft in order to prevent collisions by ensuring adequate separation between aircraft, as well as enabling the expeditious and orderly flow of air traffic \cite{icao2001doc}. In order to achieve this, the airspace is divided into multiple interlocking polyhedra of differing sizes, known as sectors, with the aircraft in each sector usually controlled by a single Air Traffic Control Officer (ATCO) with the support of a planning controller.


In order to meet future requirements, extensive ATC modernisation programmes \cite{faa,sesar} are ongoing, which seek to develop a number of advanced tools capable, for example, of providing decision support, improving workload forecasting and planning, enabling more fuel efficient procedures and systems, optimising route networks, improving network traffic predictions, and providing more accurate conflict detection tools \cite{Sesar_sol}.



A prerequisite for many of these advanced tools is an accurate Trajectory Prediction (TP), indicating the expected path an aircraft will take. Traditional TP models used within ATC are physics-based (e.g. \cite{bada_nuic}), using differential equations describing flight mechanics in order to estimate the aircraft's future positions. However, these deterministic physics-based models are insufficient for all the requirements of the modernisation initiatives, as they do not model the ATCO instructions or specific airline procedures, and, therefore, have limited utility in longer distance trajectories if these are not explicitly known. 
Current research efforts focus on TP methods that can implicitly model these factors.
\subsection{Related work}

Several previous papers have used data-driven methods to predict complete trajectories, often from the origin to the destination airports, implicitly modelling the planned routes, ATCO instructions, and procedures followed. Examples of this include the SESAR-funded DART project \cite{fernandez2017dart,sesar_dart}, which examined a range of methods including Hidden Markov Models \cite{hmm} to model origin to destination trajectories without explicitly modelling ATCO instructions. Functional representations of aircraft trajectories, derived from real-world data, have also been explored for detecting outliers \cite{nicol2013functional} and for forecasting the future trajectories of aircraft \cite{pepper2022probabilistic, TASTAMBEKOV20141}. 

Increasingly, research has focused on developing machine learning methods for the task of 4D trajectory prediction in the short- to medium-term, leveraging the large quantity of radar surveillance data available to ATC \cite{pred_analytics, N-Incept, LSTM, LSTM2, de2013machine, ZHANG_BNNs}. However, this is a challenging task due to the effects of weather uncertainties \cite{Bayes+weather, Bayes+weather2, history2}, changing intentions of ATCOs, and the effect of local procedures \cite{ZHANG2018774}. To date, such models have simplified this complexity by training machine learning models to emulate aircraft following a single route, or a limited set of routes. One solution to this issue is to use a hybrid method to `evolve' the deterministic, physics-based models that are currently used in TP. For instance, Hratsovec and Solina \cite{History} use a $k$-nearest neighbour classifier to account for the effect of airline procedures by conditioning the parameters used in BADA on the filed flight plan. Similarly, a multiple model approach has been suggested in other works (see, e.g. \cite{yepes2007new}) for modelling changing pilot intent. The generative model proposed in this work follows a similar approach by modelling the ground track followed by the aircraft, which is largely conditioned on intent, separately from the aircraft speed and climb rate, which are also affected by meteorological conditions and subject to constraints imposed by flight mechanics. 




\subsection{Contribution}

This work focuses on developing a data-driven generative model to replicate the lateral motion of aircraft through a specific sector, employing machine learning techniques to leverage the large datasets of aircraft surveillance data available to ATC, along with the planned route of the aircraft, which are filed prior to take-off. This enables prediction of the path of aircraft through this sector, implicitly taking into account the air traffic control instructions issued and sector- or route-specific procedures followed, which would otherwise be challenging to model explicitly. The generative nature of the model allows a range of realistic trajectories to be produced. By coupling these generated trajectories with existing, physics-based aircraft models such as BADA for a full 4D trajectory prediction, the model could be useful, for example, in tools for optimising arrival times, improving network flows or predicting potential traffic hotspots. 

This work evaluates the suitability of Bayesian Neural Networks (BNNs) \cite{bnn_blundell,bnn_gal,bnnreview_mackay,bnnbook_neal} and Deep Ensembles (DEs) \cite{ensemble} for this task and improves upon the works discussed in the previous section in several respects: 

\begin{enumerate}
\item Trajectories are generated in the lateral plane that are conditioned on the filed flight plan and implicitly take into account ATCO intentions 
\item A probabilistic machine learning model is trained for multiple routes in one of the busiest sectors of airspace in the UK, including converging and diverging routes 
\item As a generative model, the method is assessed by how well it can reconstruct the flow of traffic through a sector, rather than on a trajectory-by-trajectory basis
\item The method is modular and can be easily coupled with existing physics-based models, such as BADA, to give a full 4D prediction
\end{enumerate}


In the next section the piecewise linear representation used to model aircraft ground tracks is introduced and the BNN and DE architectures are outlined. Section~\ref{sec:data} describes the dataset of real-world surveillance data that was used to train and test the model. In Section~\ref{sec:application} a pair of metrics for assessing the quality of the probabilistic trajectory predictions is outlined. The performance of various probabilistic machine learning methods on the test dataset are then compared, using these metrics. 

\section{Methods}
\label{sec:methods}
The piecewise linear representation used to model aircraft ground tracks is introduced in Section~\ref{sec:lin}. Section~\ref{sec:gen} outlines the Bayesian Neural Network and Deep Ensemble methods that were used as generative models in the piecewise linear representation.


\subsection{Piecewise linear representation of aircraft ground tracks}
\label{sec:lin}
Let $\boldsymbol{x}(t)\subseteq \mathbb{X}\in \Re ^{2}$ denote an aircraft's position, at time $t$, in Geographical Coordinate Space (GCS) that passes through a sector of airspace, {denoted} $\Omega \subseteq \mathbb{X}$. The ground track of a trajectory in this airspace can be discretised as the sequence of radar observations $X=[\boldsymbol{x}^{(0)},\,\boldsymbol{x}^{(1)},\dots,\boldsymbol{x}^{(n)}]$, observed at times $\tau_0=0,\,\tau_1, \dots, \tau_n$. The distance between two points in GCS is computed using the Haversine distance, $\mathcal{D}(\boldsymbol{a},\boldsymbol{b})$ with $\boldsymbol{a}, \boldsymbol{b}\in \mathbb{X}$.

Exploiting the spatio-temporal correlation of the points in these trajectories, we define a piecewise linear basis to provide a compact representation of the trajectory. In this representation, ground tracks are piecewise linear paths that connect a set of control points $P\in[\boldsymbol{p}_0, \boldsymbol{p}_1,\dots \boldsymbol{p}_d]$, where $\boldsymbol{p}_i\subseteq{\mathbb{X}}$ refers to the location of the $i$-{th} turn. The maximum number of turns taken, $d$, is referred to as the degree of the piecewise representation. Similarly to the control points of B\'ezier curves \cite{farin_bezier}, $\boldsymbol{p}_0$ is defined to be the first point of the trajectory, i.e.,
\begin{align}
    \boldsymbol{p}_0 :=\boldsymbol{x}^{(0)}.
\end{align}
To account for the unequal lengths of the trajectories in the training data, each trajectory is interpolated onto a normalised timescale $t\in[0,1]$, with:
\begin{align}
    \boldsymbol{x}(t_0)=\boldsymbol{p}_0 \;\text{and}\; \boldsymbol{x}(t_d)=\boldsymbol{p}_d,
\end{align}
where $t_0=0$ and $t_d=1$. Arrival times $t_i,\, i=1,\dots d-1$ at the control points collected in $P$, using the normalised timescale, are proportional to the length of the trajectory travelled:
\begin{align}
    t_i=t_{i-1}+\frac{\mathcal{D}(\boldsymbol{p}_i,\boldsymbol{p}_{i-1})}{\sum_{i=1}^d \mathcal{D}(\boldsymbol{p}_i,\boldsymbol{p}_{i-1})}, \; i=1:d-1.
    \label{eq:rep_times}
\end{align}
Using the arrival times and control points in $P$, the GCS coordinates of a point in the piecewise linear representation can be determined through:
\begin{align}
    \boldsymbol{x}(t)=\boldsymbol{x}(t_i)+\frac{t-t_i}{t_{i+1}-t_i}({\boldsymbol{p}}_{i+1}-{\boldsymbol{p}}_i),
    \label{eq:rep_x}
\end{align}
where ${t}_{i}\leq t \leq {t}_{i+1}$, with $t_{i}$ referring to the time at which the last control point in $P$ was reached and $t_{i+1}$ to the arrival time at the next control point. For an individual trajectory, $X$, in the training data, an optimisation process was followed to determine the optimal locations of the points in $P$ that minimised a cost function $\mathcal{J}(\cdot)$:
\begin{align}
        \hat{P}=\underset{P}{\textrm{arg min}}\;\mathcal{J}(P|X).
\end{align}
$\mathcal{J}(\cdot)$ quantifies how well the piecewise linear representation fits the radar data in $X$, and is defined as:
\begin{align}
    \mathcal{J}(P)=\sum_{j=0}^n\Big|\Big|\boldsymbol{x}^{(j)}-\hat{\boldsymbol{x}}^{(j)}\Big|\Big|^2+\lambda\mathcal{I}(\hat{X}), 
    \label{eq:opt}
\end{align}
where $\hat{\boldsymbol{x}}^{(j)}$ refers to a point at time $t=\tau^{(j)}/\tau^{(n)}$ along the trajectory generated by \eqref{eq:rep_times} and \eqref{eq:rep_x}, conditioned on $P$, and collected as $\hat{X}$. The cost function consists of two components, a data fit term in the form of the  $L2$ distance between $X$ and $\hat{X}$, and a penalty term to dissuade trajectories from making many large turns.
The indicator function, $\mathcal{I}$ is defined as:
\begin{align}
    \mathcal{I}(\hat{X})=\begin{cases}
0\;\;\text{if}\;\phi_{\Undersig}(\hat{X})\leq \phi_u \\
1\;\;\text{otherwise}
    \end{cases}\!\!\!\!,
\end{align}
where $\phi_{\Undersig}(\hat{X})$ represents the total angle turned through by the trajectory and $\phi_u$ is a user-selected maximum bound on this quantity.
The regularisation constant, $\lambda$, is set to a value large enough that if the total angle turned by the trajectory exceeds $\phi_u$ a severe penalty is given. $\lambda$ was adjusted to be an order of magnitude greater than the typical $L2$ distance between $X$ and $\hat{X}$ in the training data. Similarly, $\phi_u$ was set to the largest $\phi_{\Undersig}$ observed in the training data.

\subsection{Generative model for aircraft ground tracks}
\label{sec:gen}
This section presents a sector-specific generative model for aircraft ground tracks based on an aircraft's filed flight plan and its state upon entering.
This information is collected in the vector $\boldsymbol{\xi}$. The model generates ground tracks by performing the probabilistic mapping $\boldsymbol{\xi}\rightarrow \{\boldsymbol{p_i}\}_{i=1}^d$, where $P=\{\boldsymbol{p_i}\}_{i=0}^d$ contains a sequence of control points in the piecewise linear representation discussed above. Here, $\boldsymbol{p}_0$ represents the GCS coordinates at which the aircraft enters $\Omega$ and anchors the generated ground track to the aircraft's initial location. The conditional distribution $P|\boldsymbol{\xi}$ can be used to generate the ground track using \eqref{eq:rep_times} and \eqref{eq:rep_x}.
The quantities collected in $\boldsymbol{\xi}$ are:
\begin{itemize}
    \item The aircraft position $\boldsymbol{p}_0$ in GCS coordinates on entering ${\Omega}$ 
    \item The aircraft flight level (altitude) when entering $\Omega$
    \item The filed flight plan (encoded as a sequence of GCS coordinates)
\end{itemize}

$\boldsymbol{\xi}$ necessarily only contains partial information about the context of the trajectory. The strategy of the ATCO for handling the traffic in their sector will also affect the trajectory of the aircraft. This will, in part, be determined by factors such as the time of day, the number of aircraft in the sector, and the state of neighbouring sectors. It is, therefore, difficult to encode the strategy of the ATCO directly in the model. For this reason, probabilistic machine learning is used to inject uncertainty into the model predictions that accounts for this epistemic uncertainty. In this paper the suitability of Deep Ensembles \cite{ensemble} and Bayesian Neural Networks \cite{bnn_blundell,bnn_gal,bnnreview_mackay,bnnbook_neal} for this task are investigated. Both methods are popular ways of adding uncertainty to neural network predictions in the literature \cite{bayesiandl_wilson}. We outline the main features of the methods below. For convenience, we denote the vector that collects the set $\{\boldsymbol{p_i}\}_{i=1}^d$ as $\boldsymbol{y}\in\Re^{2d}$. Evaluating \eqref{eq:opt} for every trajectory in a dataset of $n_f$ individual flights through $\Omega$ yields the training set $D=\{\boldsymbol{\xi}^{(k)},\boldsymbol{y}^{(k)}\},\,k=1:n_f$. \newline
\newline
\noindent \textbf{Deep Ensembles (DEs)}
\newline
Lakshminarayanan et al. \cite{ensemble} proposed to use an ensemble of $n_e$ independently-trained neural networks as a way to estimate predictive uncertainty using neural networks. Each of the $n_e$ networks consist of the same architecture, but are initialised with different weights $\boldsymbol{\theta}\in\Re^{n_{{\theta}}}$. Once trained, the mean and standard deviation of the networks' outputs are used as the parameters of i.i.d. Gaussian distributions that represent the ensemble's average prediction and associated uncertainty. In this work, we model each component of $\boldsymbol{y}$ as the output of the deep ensemble. Each network is trained independently to minimise the loss function:
\begin{align}
   \hat{\boldsymbol{\theta}}=\underset{\boldsymbol{\theta}}{\text{arg min}}\;\mathcal{L}(D|\boldsymbol{\theta}),
   \label{eq:nn1}
\end{align}
where $\mathcal{L}(\cdot)$ is defined as:
\begin{align}
    {\mathcal{L}(D|\boldsymbol{\theta})=r(\boldsymbol{\theta})+\sum_{k=1}^{n_f}l\big(f(\boldsymbol{\xi}^{(k)}|\boldsymbol{\theta}) , \boldsymbol{y}^{(k)}\big)},
    \label{eq:nn2}
\end{align}
and where $r(\cdot)$ is a regularisation term, $f(\cdot)$ represents the output of the network, and $l(\cdot)$ is the negative-log likelihood. Having trained the $n_e$ networks in the ensemble, using the Adam optimiser \cite{adam_kingma}, the predictions of the ensemble for an unseen context, $\boldsymbol{\xi}$, are represented as the set of i.i.d. Gaussian probability distributions $\boldsymbol{y}_m=N(\mu^*_m,\sigma^*_m), \,m=1:2d$, with:
\begin{align}
    \mu^*_m(\boldsymbol{\xi})&=\frac{1}{n_e}  \sum_{o=1}^{n_e} \mu_{o,m}(\boldsymbol{\xi}),\\
    \sigma^{*2}_m(\boldsymbol{\xi})
    &= {\sigma^{2}_{\mu_m}(\boldsymbol{\xi})+\sigma^{2}_{\sigma_m}(\boldsymbol{\xi})}, \nonumber 
\end{align}
where the standard deviation of the ensemble contains contributions from the variance of the individual networks, $\sigma^*_\sigma$, and the spread of their means, $\sigma^*_\mu$, with:
\begin{align}
    \sigma^{2}_{\mu_m}(\boldsymbol{\xi})&= \frac{1}{n_e} \sum_{o=1}^{n_e} \mu^2_{o,m}(\boldsymbol{\xi})-\mu_m^{*2}(\boldsymbol{\xi}), \\
    \sigma^{2}_{\sigma_m}(\boldsymbol{\xi})&=\frac{1}{n_e}\sum_{o=1}^{n_e}\sigma_{o,m}^2(\boldsymbol{\xi}). \nonumber
\end{align}
The ensemble can be used to generate trajectories by sampling the coordinates of the control points from the i.i.d. Gaussian distributions, and using \eqref{eq:rep_times} and \eqref{eq:rep_x} to determine the ground track that connects these control points.
\newline
\newline
\noindent \textbf{Bayesian Neural Networks (BNNs)}
\newline 
Deep ensembles incorporate uncertainty into neural network predictions by expressing the outputs as Gaussian distributions, rather than as point estimates. A drawback of this approach is that each element in posterior samples of $\boldsymbol{y}$ are sampled independently. An alternative approach is to add uncertainty by making all \cite{bnnbook_neal} or a subset \cite{daxberger2021laplace} of the network parameters, $\boldsymbol{\theta}$, probabilistic. Bayesian methods are then used to infer these probability distributions, based on the evidence of the training data. The advantage of this formulation is that it provides a richer probabilistic model for $\boldsymbol{y}$, because the outputs are no longer constrained to be i.i.d. Gaussian distributions. On the other hand, BNN training is more complex compared to deep ensembles because standard backpropagation methods cannot be applied \cite{jospin2022hands}. Methods such as variational inference are commonly criticised for being challenging to implement and expensive to train \cite{osawa2019practical}. For this reason, in this work, the posteriors of the uncertain network parameters were estimated using the Laplace approximation \cite{daxberger2021laplace}. This is a two-step process: first a \emph{maximum a posteriori} (MAP) estimate is generated for $\boldsymbol{\theta}$ by solving \eqref{eq:nn1} and \eqref{eq:nn2}, with the $L2$ distance used for $l(\cdot)$. Having obtained the MAP estimate, denoted $\boldsymbol{\theta}_{MAP}$, the posterior of the uncertain parameters is locally approximated with a Gaussian distribution. This distribution is centred at $\boldsymbol{\theta}_{MAP}$, with its covariance matrix corresponding to the local curvature:
\begin{align}
    p(\boldsymbol{\theta}|D)=N(\boldsymbol{\theta}|\boldsymbol{\theta}_{MAP},\Sigma)\;\text{with}\;\Sigma:=(\nabla_{\boldsymbol{\theta}}^2\mathcal{L}(D|\boldsymbol{\theta})_{\boldsymbol{\theta}_{MAP}})^{-1},
\end{align}
and where $p(\cdot)$ represents the posterior probability of the network parameters. The Laplace package \cite{daxberger2021laplace} was used to implement this post-hoc step, using a zero-mean Gaussian prior for the uncertain network parameters. The generalized Gauss-Newton matrix (GGN) was computed to store the derivatives of the network parameters. Two strategies were deployed to reduce the computational cost of this step: firstly, by only considering a subset of the network parameters to be uncertain and secondly, by employing factorization assumptions for the GGN matrix. Following an initial exploratory phase, the best results were found using a diagonal factorization with a last-layer Laplace approximation. 
\newline
\newline
\noindent \textbf{Linear model}
\newline
A probabilistic linear model is used as a baseline with which to compare the machine learning methods to. For lateral TP, a first assumption is that aircraft will fly directly to the exit of the sector. In the linear model used here, this exit point is the last waypoint in the flight plan. To make the model probabilistic, jitter was added to the GCS coordinates of the exit point, using a two-dimensional, zero-mean, Gaussian distribution with covariance $\Sigma=\sigma^2 I$, where $\sigma=0.05$\textdegree.
\newpage
\section{Data Preparation}
\label{sec:data}
\begin{figure}
    \centering
    \includegraphics[width=0.6\textwidth]{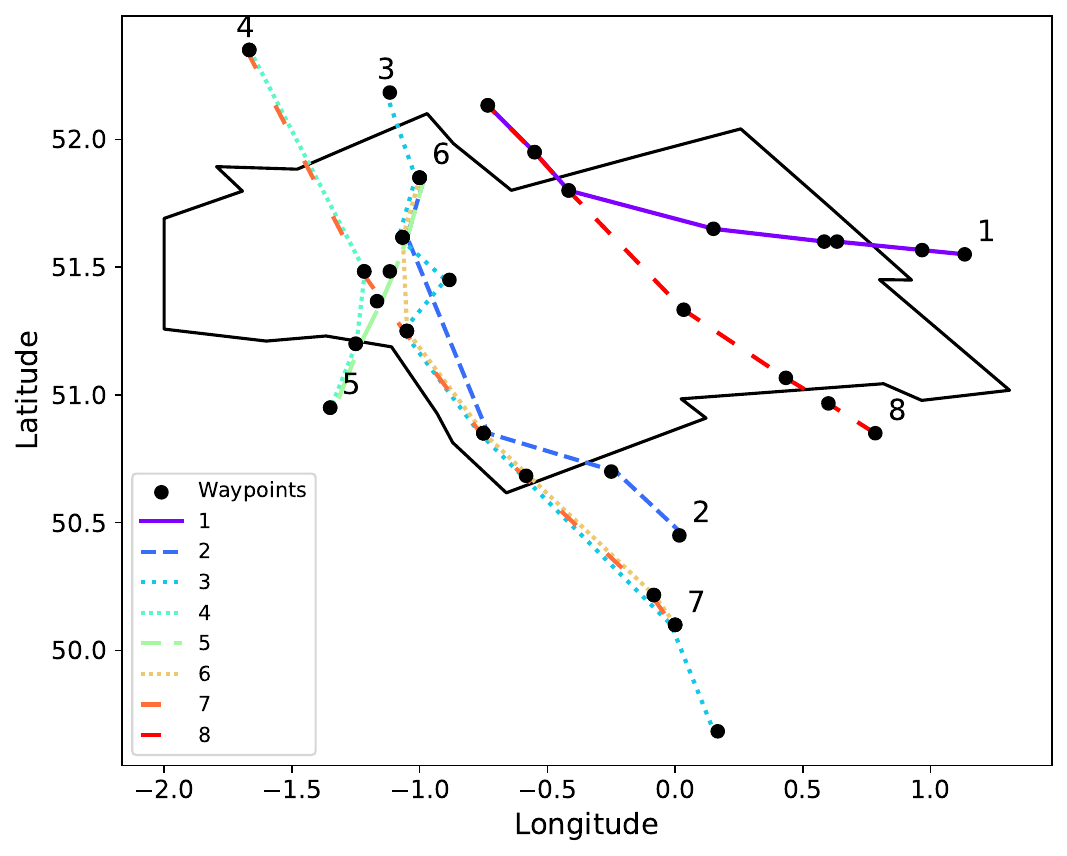}
    \includegraphics[width=0.39\textwidth]{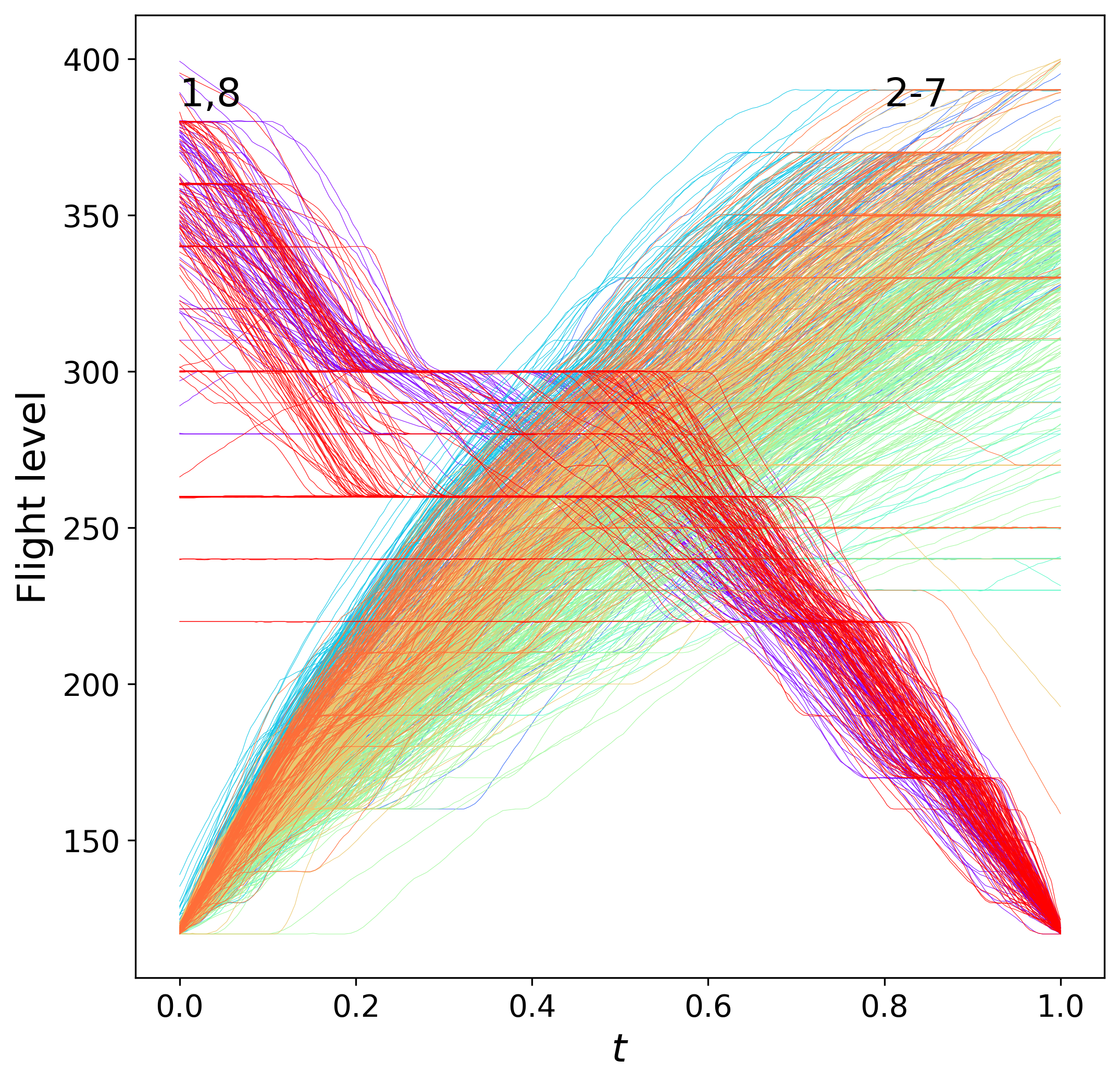}
    \includegraphics[width=\textwidth]{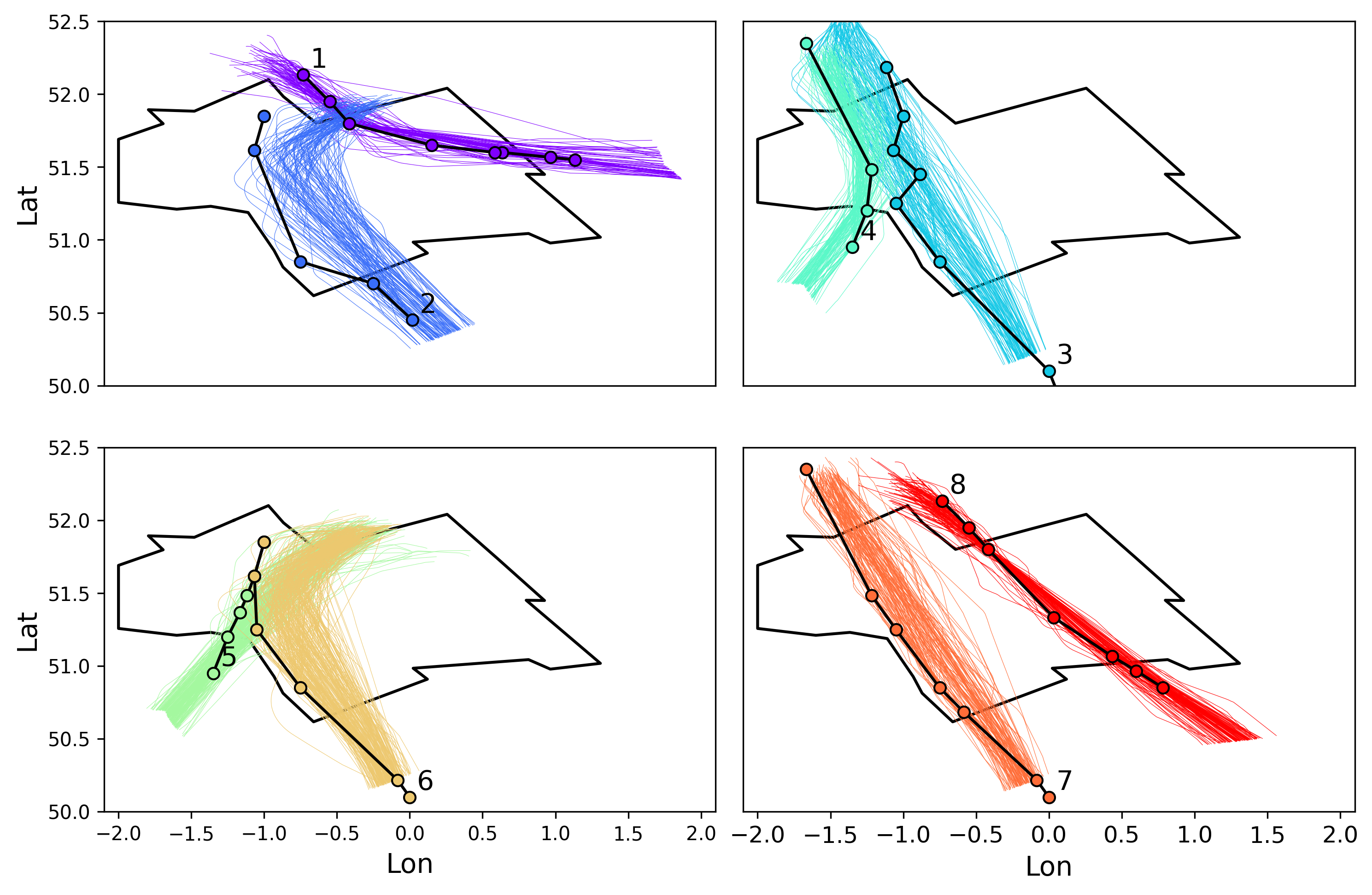}
    \caption{Boundary of the studied sector, locations of the waypoints, and the eight unique routes through $\Omega$ in the dataset (top-left). Flight levels of the trajectories in the dataset, interpolated onto $t\in[0,1]$ (top-right). Trajectories in the dataset, coloured by their flight plan through $\Omega$ (bottom).} 
    \label{fig:data_rte} 
\end{figure}

Probabilistic machine learning models were trained to emulate the flow of air traffic through a specific sector located in the upper airspace of the UK, above several of the world's busiest airports. $\Omega$ was defined to include the sector and a 50 nautical mile buffer around it's edge. The sector boundary in the lateral plane is illustrated in Fig.~\ref{fig:data_rte}. The dataset contained one week's worth of radar surveillance data, filtered to the 1902 aircraft that were under the control of ATCOs assigned to this sector during this period. For ease of comparison, another filtering step was performed: selecting those flights that followed the eight most popular filed flight plans in the dataset. Of the 1152 remaining trajectories, an 80:20 training:test split was performed, resulting in a training dataset of 922 flights and a test dataset of 230 flights. The top-left panel of Fig.~\ref{fig:data_rte} displays the GCS coordinates of the waypoints in the filed routes, together with lines connecting the waypoints of each unique flight plan.

The ground tracks of these trajectories are illustrated in the lower panels of Fig.~\ref{fig:data_rte}, with lines coloured by the filed flight plan. As can be seen from Fig.~\ref{fig:data_rte}, aircraft are not constrained to fly from one fix to the next and there can be considerable variation between the route joining together the waypoints and the actual trajectory followed. Many of these routes are overlapping, with some routes sharing waypoints. Flows of traffic in the sector are dominated by aircraft either departing from or arriving at major airports in the south of the UK. The top-right panel of Fig.~\ref{fig:data_rte} indicates the flight level (altitude) of the aircraft on the normalised timescale. Routes 1 and 8 correspond to descending aircraft entering the sector from the east and south, respectively, and heading north. Routes 2-7 correspond to climbing aircraft that enter the sector from the north, having departed from these airports, and are heading south. 

The trajectories in the training dataset were projected onto the piecewise linear basis discussed in Section~\ref{sec:methods}, with the control points determined by solving \eqref{eq:opt} for each trajectory. Based on an inspection of the typical number of turns taken by aircraft in this dataset, the degree of the representation was set to $d=3$.

\newpage
\section{Application to a sector of UK airspace}
\label{sec:application}
The probabilistic machine learning models described in Section~\ref{sec:methods} were trained using the training set outlined in the previous section, and, for each of the test trajectories, $\boldsymbol{\xi}$, 10 posterior samples were drawn. In order to assess how well the methods emulated the flow of traffic, it was necessary to quantify the difference between the generated trajectories and the trajectories in the test dataset. While there are numerous methods in the literature for computing the distance between trajectories (see, e.g. \cite{traj_sim}), there is no fixed convention for measuring trajectory similarity in ATC. Common metrics for measuring trajectory similarity such as the Euclidean distance, Hausdorff distance~\cite{hausdorff_taha}, and Dynamic Time Warping~\cite{dtw_berndt} implicitly account for differences in heading. However, in ATC the relative heading between trajectories is also an important consideration. As an example, trajectories in routes 3 and 4 overlap for much of the sector, but, from an ATC perspective, would be considered to be distinct because they exit the sector across different boundaries. Summarising the similarity of trajectories with a single statistic would require finding a weighting between the relative headings of trajectories and their spatial proximity for ATC. For this reason, we simplify the measurement of trajectory similarity by assessing how well the probabilistic methods emulate the flow of air traffic through a sector using a geometric method. The advantage of this approach is that the spatial proximity and relative headings between two sets of trajectories can be quantified separately.

\begin{figure}
    \centering
    \includegraphics[width=\textwidth]{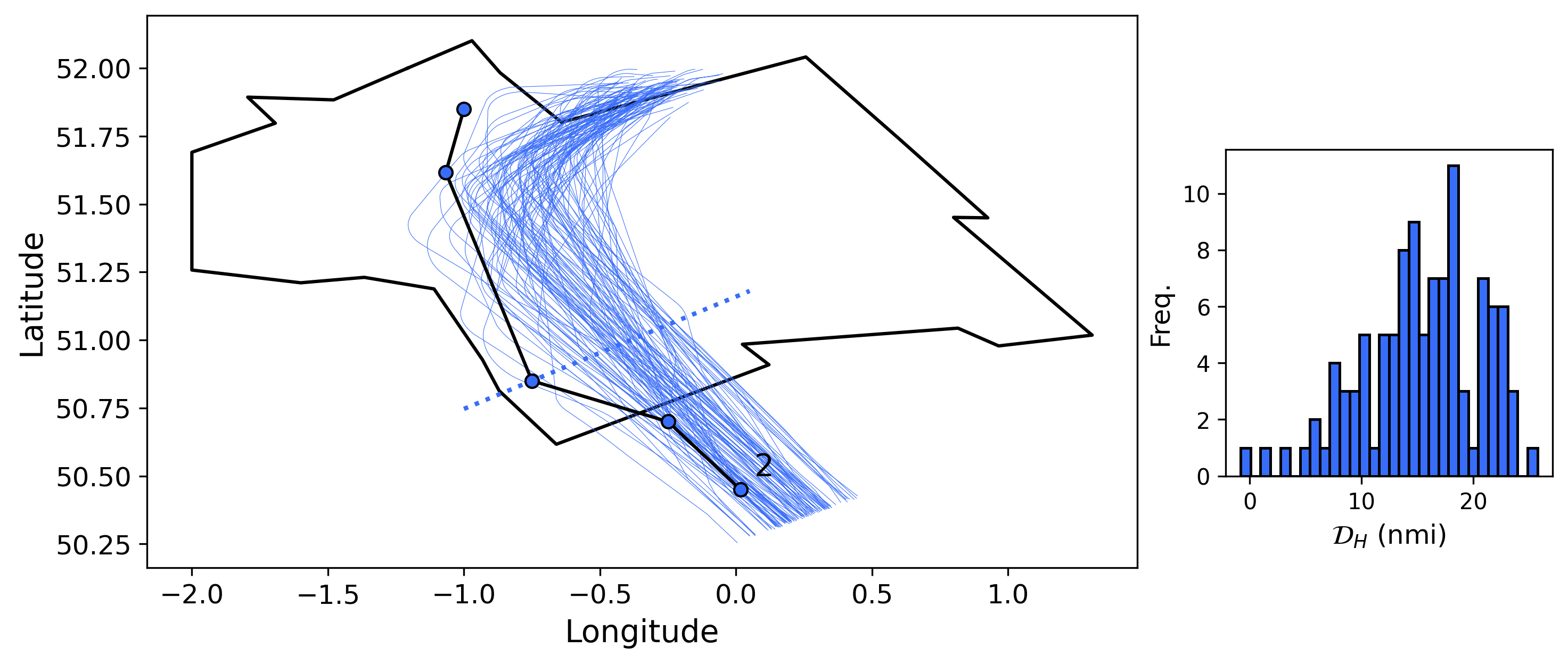}
    \caption{Schematic illustrating the geometric method used to assess how well the proposed generative emulated the distribution of ground tracks in the dataset.
    }
    \label{fig:assess}
\end{figure}

Fig.~\ref{fig:assess} is a schematic that illustrates the method using the training data. For a given route, in this case route 2, a plane is defined that is perpendicular to the filed flight plan. The origin of this plane is the last waypoint that is within the sector boundary. This plane is represented by a dashed blue line in Fig.~\ref{fig:assess}. The Haversine distance, in nautical miles, between the origin and the point at which a trajectory intersects this plane is calculated, along with the heading of the trajectory at this point. These quantities are denoted $\mathcal{D}_H$ and $\phi$ respectively. Histograms of $\mathcal{D}_H$ and $\text{sin}(\phi)$ can then be computed for each of the eight studied routes. The sine of the heading is used in order to make the headings continuous. The schematic in the right panel of Fig.~\ref{fig:assess} illustrates the histogram extracted for $\mathcal{D}_H$ from the training data for route 2. For a given route, the similarity of two sets of trajectories is assessed by comparing the statistical distance between the histograms of these two quantities.

\begin{figure}
    \centering
    \includegraphics[width=\textwidth]{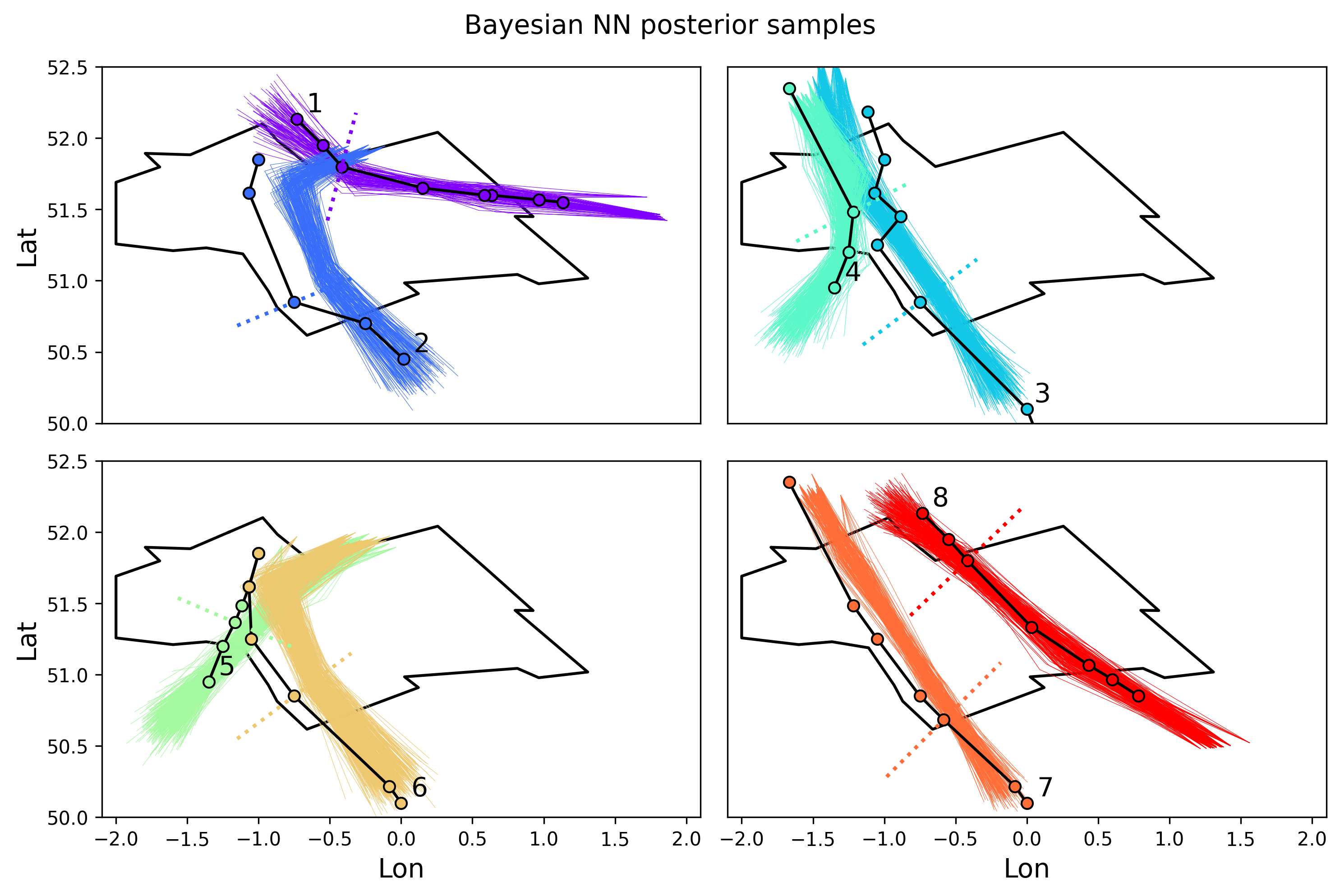}
    \includegraphics[width=\textwidth]{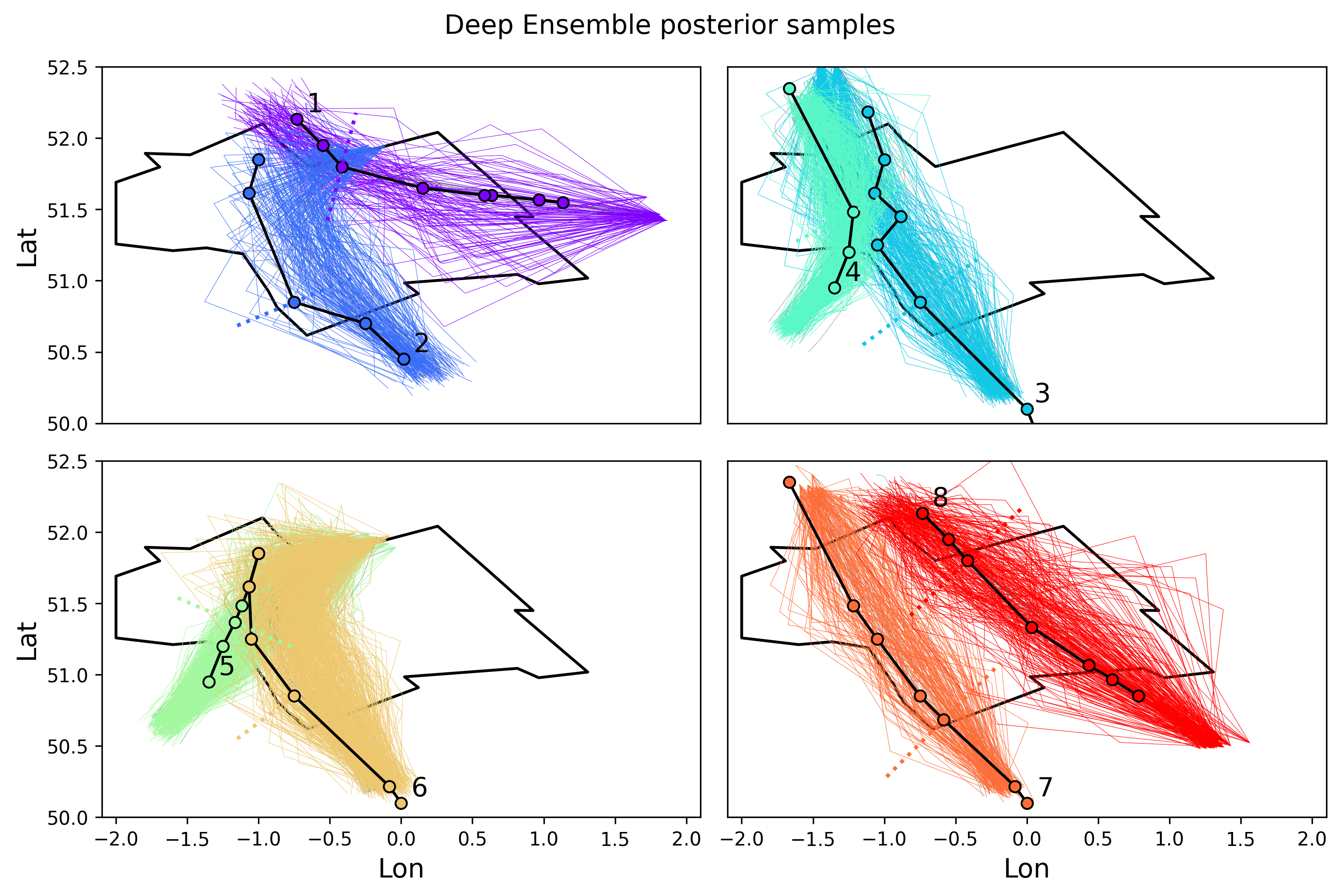}
    \caption{Generated trajectories, sampled from the posteriors of the Bayesian Neural Network (top) and Deep Ensemble (bottom). Solid black lines indicate the waypoints of the route, with dotted lines indicating the plane on which $\mathcal{D}_H$ and $\phi$ are computed for each route.}
    \label{fig:post}
\end{figure}

\begin{figure}
    \centering
    \includegraphics[width=\textwidth]{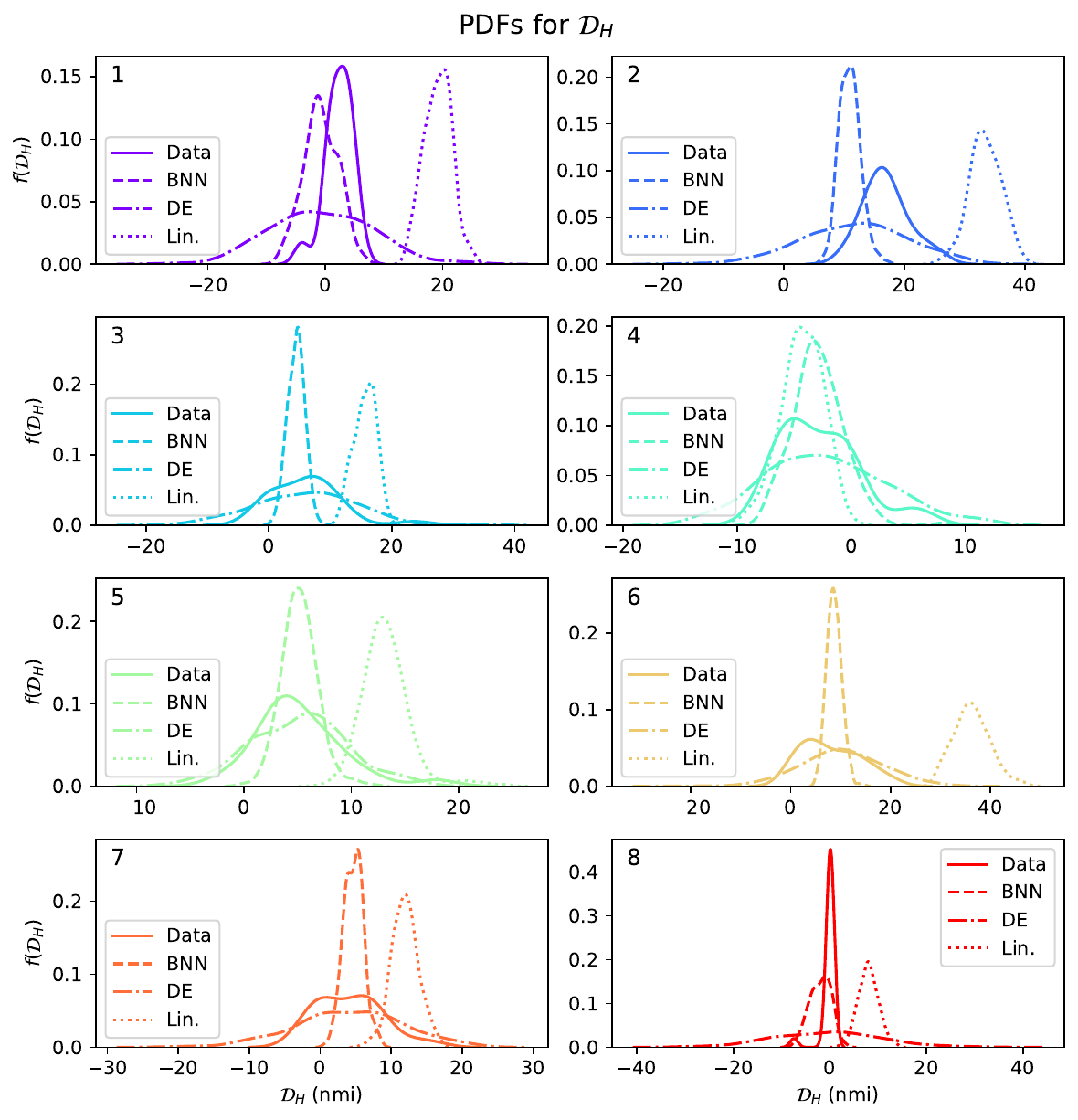}
    \caption{Kernel density plot of $\mathcal{D}_H$, coloured by unique route through $\Omega$, for the test dataset and the trajectories generated by the probabilistic models.}
    \label{fig:kde}
\end{figure}

\begin{figure}
    \centering
    \includegraphics[width=\textwidth]{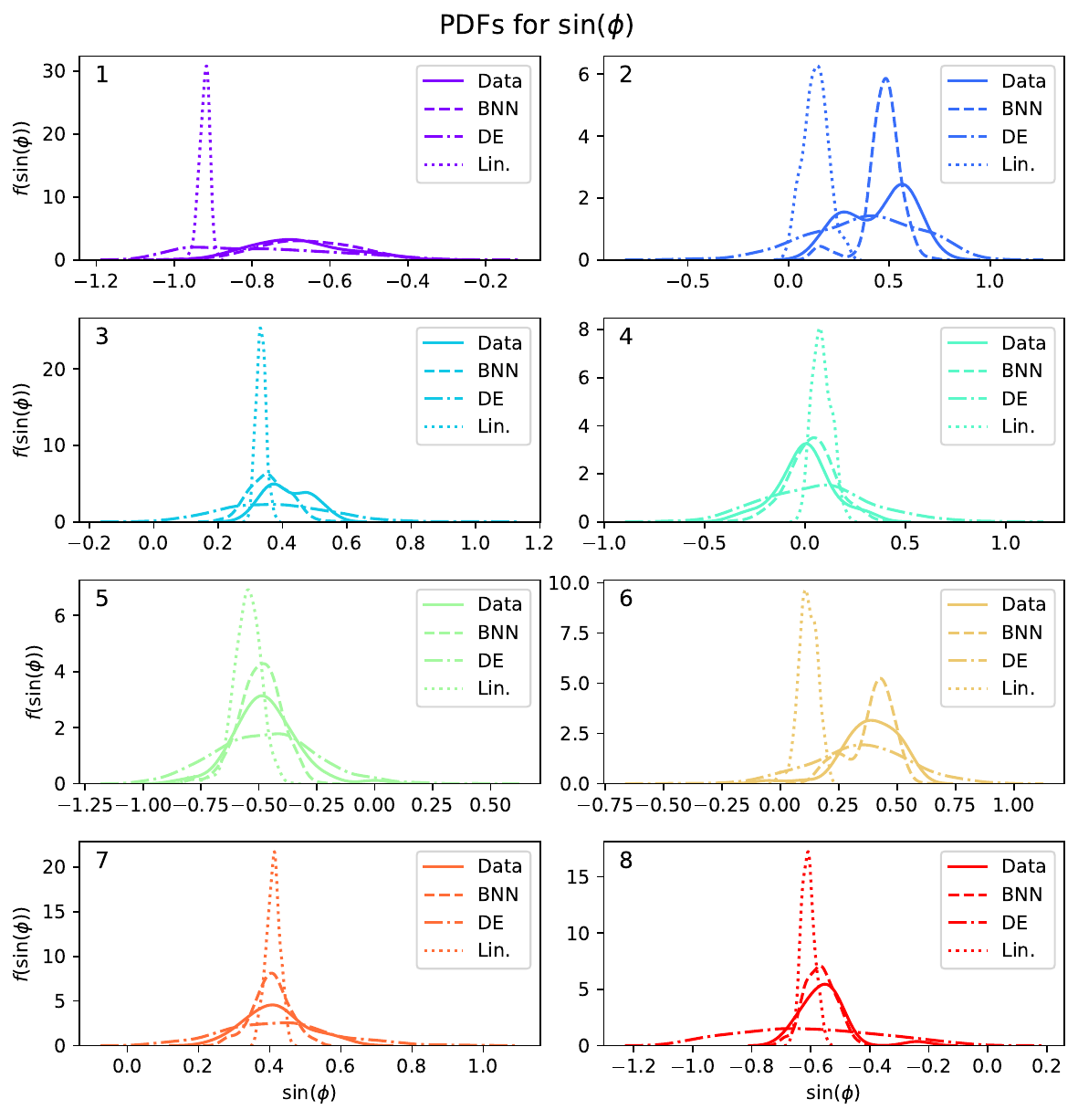}
    \caption{Kernel density plot of $\text{sin}(\phi)$, coloured by unique route through $\Omega$, for the test dataset and the trajectories generated by the probabilistic models.}
    \label{fig:kde_head}
\end{figure}

Fig.~\ref{fig:post} illustrates the trajectories generated by the BNN (top) and the Deep Ensemble (bottom). The plots also show the waypoints of each route as coloured dots, joined by solid lines. The dashed coloured lines indicate the planes that were used to compute $\mathcal{D}_H$ and $\phi$. The posterior samples from the Deep Ensemble are more widely distributed across the sector, however, the consequence of this is that many samples make turns that are not observed in the surveillance data. Conversely, the posterior samples from the BNN are more tightly clustered, but, from inspection, appear both more similar to the trajectories in Fig.~\ref{fig:data_rte}, and more physically plausible. Fig.~\ref{fig:kde} displays kernel density estimates for the Probability Density Functions (PDFs) of $\mathcal{D}_H$ for the test dataset, the BNN, Deep Ensemble, and the probabilistic linear model. Similarly, Fig.~\ref{fig:kde_head} displays kernel density estimates of the PDFs for $\text{sin}(\phi)$ for the test data and the various methods. Tables~\ref{tab:KS}~and~\ref{tab:mean} tabulate the statistical distance between the probability distributions of the data and the three methods, as quantified by the Kolmogorov-Smirnov \cite{kstest_massey} (KS) distance and percentage difference of the mean respectively. The statistical distance for the best performing model is in bold for both quantities and each of the eight routes.

As might be expected, the linear method performed best for route 8, where the majority of trajectories are straight lines across the sector, but it does not perform well for the other routes which have more complex trajectories. In general, the distribution of $\mathcal{D}_H$ for the Deep Ensemble is closer to that of the test data than the BNN. The distributions from the BNN tend to accurately estimate the mean but, except for the first and eighth routes, it underestimates the variance. However, the headings at the plane intersection of the trajectories generated by the BNN are much closer to the test dataset than the other two generative models. This is likely due to the fact that samples generated from the BNN have correlations between elements in $\boldsymbol{y}$. In contrast, the samples for the $2d$ elements of $\boldsymbol{y}$ from the DE are independent, and, therefore, often introduce spurious turns into the generated trajectories; c.f. route 8 in Fig.~\ref{fig:post}.

\setlength{\tabcolsep}{1em}

\begin{table}[]
\centering
\caption{Statistical distance between distributions for $\mathcal{D}_H$ and $\text{sin}(\phi)$, as quantified by the KS distance, for each of the eight unique routes. }
\label{tab:KS}
\begin{tabular}{*1c*3c|*3c}
\toprule
Route & \multicolumn{3}{c}{KS distance ($\mathcal{D}_H$) $\downarrow$} &\multicolumn{3}{c}{KS distance ($\text{sin}(\phi)$) $\downarrow$}\\
& Lin. & DE & BNN& Lin. & DE & BNN \\
\midrule
1 & 1.000 & \textbf{0.483} & 0.578 & 1.000 & 0.378  & \textbf{0.178} \\
2 & 0.905 & \textbf{0.457} & 0.805 & 0.848 & \textbf{0.267} & 0.352 \\
3 & 0.648 & \textbf{0.188} & 0.472 & 0.772 & \textbf{0.344} & 0.356\\
4 & 0.912 & \textbf{0.235} & 0.238 & 0.527 & 0.262 & \textbf{0.142}\\
5 & 0.553 & \textbf{0.138} & 0.233 & 0.420 & 0.171 & \textbf{0.111}\\
6 & 1.000 & \textbf{0.200} & 0.464 & 0.966 & 0.205 & \textbf{0.150}\\
7 & 0.482 & \textbf{0.206} & 0.406 & 0.341 & 0.247 & \textbf{0.176}\\
8 & 0.908 & \textbf{0.479} & 0.692 & 0.562 & 0.404 & \textbf{0.179}\\
\bottomrule
\end{tabular}
\end{table}

\begin{table}[]
\centering
\caption{Statistical distance between distributions for $\mathcal{D}_H$ and $\text{sin}(\phi)$, as quantified by the percentage error in the mean, for each of the eight unique routes.}
\label{tab:mean}
\begin{tabular}{*1c*3r|*3r}
\toprule
Route & \multicolumn{3}{c}{$\Delta \mathbb{E}(\mathcal{D}_H$) (\%) $\downarrow$} &\multicolumn{3}{c}{$\Delta\mathbb{E}(\text{sin}(\phi))$ (\%) $\downarrow$}\\
& Lin. & DE & BNN& Lin. & DE & BNN \\
\midrule
1 &  506.45 & \textbf{122.89} & 129.61 & 55.43 & 26.16 & \textbf{2.7}\\
2 &   67.48 & \textbf{30.08} & 35.47 & 10.47 & 18.74 & \textbf{0.58}\\
3 &   74.58 & \textbf{5.04} & 27.19 & 3.44 & \textbf{0.42} & 2.38\\
4 &  238.51 & 19.65 & \textbf{5.11} & 1121.90 & \textbf{134.94} & 227.64\\
5 &   58.20 & 1.06 & \textbf{0.37} & 1.03 & \textbf{0.02} & 3.64\\
6 &  289.74 & 27.19 & \textbf{8.45} & 13.56 & 9.33 & \textbf{5.43}\\
7 &   68.97 & \textbf{1.17} & 14.23 & \textbf{0.44} & 0.47 & 0.64\\
8 & 1352.92 & \textbf{132.48} & 739.97 & 11.17 & 19.10 & \textbf{2.80}\\
\bottomrule
\end{tabular}
\end{table}

\FloatBarrier
\section{Conclusions}
\label{sec:conclusion}
A probabilistic machine learning method has been presented for generating plausible ground tracks for an aircraft entering a specific sector of airspace. The model is conditioned on contextual information gathered from real-world aircraft surveillance data. A piecewise linear representation was used to model the ground tracks. It was found that, of the probabilistic models evaluated, the best performance was achieved using a Bayesian Neural Network, with posterior probability distributions over the weights in the last-layer provided by the Laplace Approximation. This is likely due to the natural way in which correlations between the locations of control points can be encoded within the model. It was found that this probabilistic model could be used to plausibly emulate the flow of air traffic through a busy sector of UK airspace with real-world data. 

Future work includes coupling the presented probabilistic models for aircraft ground tracks with a deterministic, physics-based TP model such as the Base of Aircraft Data (BADA) model \cite{bada_nuic}. Combined, these models would provide a probabilistic method for four-dimensional TP (GCS coordinates, time, and altitude) that would be capable of propagating the effect of epistemic uncertainty arising from unknown pilot and ATCO intentions. 

\bibliographystyle{IEEEtran}
\bibliography{IEEEabrv,references}

\end{document}